\documentstyle[12pt]{article}
\newcommand{\beq}{\begin{equation}}
\newcommand{\eeq}{\end{equation}}
\newcommand{\bdm}{\begin{displaymath}}
\newcommand{\edm}{\end{displaymath}}
\newcommand{\beqr}{\begin{eqnarray}}
\newcommand{\eeqr}{\end{eqnarray}}
\begin{document}
\title{On the derivatives of generalized Gegenbauer polynomials}
\author{W. Garc\'{\i}a Fuertes, A. M. Perelomov\footnote{On leave of absence from the Institute for Theoretical and Experimental Physics, 117259, Moscow, Russia. Current E-mail address: perelomo@dftuz.unizar.es}\\ {\small\em Departamento de F\'{\i}sica, Facultad de Ciencias, Universidad de
Oviedo, E-33007 Oviedo, Spain}} 
\date{}
\maketitle
\begin{abstract}
We prove some new formulae for the derivatives of the generalized Gegenbauer polynomials associated to the Lie algebra $A_2$.
\end{abstract}
\section*{}
As it is well known \cite{grad}, the classical Gegenbauer polynomials $C_m^\kappa(z)$ suffer, when differentiated in $z$, a shift in the parameter $\kappa$, namely
\beqr
\frac{d P_m^\kappa}{d z}&=&m\, P_{m-1}^{\kappa+1},\nonumber\\
P_m^\kappa(z)&=&\frac{m!}{(\kappa)_m}C_m^\kappa(\frac{z}{2}),\ \ \ \ \ \ (\kappa)_m=\kappa (\kappa+1)\ldots(\kappa+m-1)\nonumber
\eeqr
The classical Gegenbauer polynomials are (up to a factor) the eigenfunctions of the simplest quantum Calogero-Sutherland Hamiltonian \cite{ca71},\cite{su72}, that related to the Lie algebra $A_1$. It is the purpose of this note to show that the same shift in $\kappa$ takes place in the derivatives of the generalized Gegenbauer polynomials $P_{m,n}^\kappa (z_1,z_2)$ giving the quantum eigenfuctions of the Calogero-Sutherland system with Lie algebra $A_2$:
\beqr
\Delta^\kappa P_{m,n}^\kappa&=&\varepsilon_{m,n}(\kappa)\,P_{m,n}^\kappa(z_1,z_2),\nonumber\\
P_{m,n}^\kappa&=&z_1^m z_2^n+{\rm lower\ terms},\nonumber\\
\Delta^\kappa&=&(z_1^2-3z_2)\partial_{z_1}^2+(z_2^2-3z_1)\partial_{z_2}^2+(z_1 z_2-9)\partial_{z_1}\partial_{z_2}+(3\kappa+1)(z_1\partial_{z_1}+z_2\partial_{z_2})\nonumber\\
\varepsilon_{m,n}(\kappa)&=&m^2+n^2+mn+3\kappa(m+n),\nonumber
\eeqr
see \cite{op83},\cite{pe98a},\cite{prz98},\cite{pe99},\cite{pe00}. Specifically, we will prove the following formulae:
\beqr
\frac{\partial P_{m,n}^\kappa}{\partial z_1}&=&m\, P_{m-1,n}^{\kappa+1}+A_{m,n}(\kappa)\, P_{m-2,n-1}^{\kappa+1}+B_{m,n}(\kappa)\, P_{m,n-2}^{\kappa+1}\label{formula1}\\
\frac{\partial P_{m,n}^\kappa}{\partial z_2}&=&n\, P_{m,n-1}^{\kappa+1}+A_{n,m}(\kappa)\, P_{m-1,n-2}^{\kappa+1}+B_{n,m}(\kappa)\, P_{m-2,n}^{\kappa+1},\label{formula2}
\eeqr
where
\beqr
A_{m, n}(\kappa)&=& \frac{m (m - 1) n (m + n + \kappa - 1) (m + n + \kappa)}{(m + \kappa - 1) (m + \kappa) (m + n + 2 \kappa - 1) (m + n + 2 \kappa)}\nonumber\\
B_{m, n}(\kappa)&=&-\frac{n (n - 1) (m + n + \kappa)}{(n + \kappa - 1) (n + \kappa)}\label{coef1}.
\eeqr
Consider first (\ref{formula1}). The proof of this formula proceeds by induction on the second quantum number. The generating function for the Jack polynomials $P_{m,0}^\kappa$ is known to be \cite{ja70}
\beq
(1-z_1 t+z_2 t^2-t^3)^{-\kappa}=\sum_{m=0}^\infty \frac{(\kappa)_m}{m!}P_{m,0}^\kappa t^m.
\eeq
Differentiation of this expression shows the validity of (\ref{formula1}) when $n=0$. On the other hand, we can use the recurrence relations for the generalized Gegenbauer polynomials \cite{pe99} to express $P_{m,n}^\kappa$ in terms of polynomials with lower $n$:
\beq
P_{m,n}^\kappa=z_2\, P_{m,n-1}-
\tilde{a}_{m,n-1}(\kappa)\, P_{m-1,n-1}^\kappa-c_{n-1}(\kappa)\, P_{m+1,n-2}^\kappa\label{rec}
\eeq
with
\beqr
\tilde{a}_{m,n}(\kappa)&=&\frac{m(n+m+\kappa)(m-1+2\kappa)(n+m-1+3\kappa)}{(m+\kappa)(m-1+\kappa)(n+m+2\kappa)(n+m-1+2\kappa)},\nonumber\\
c_{n}(\kappa)&=&\frac{n(n-1+2\kappa)}{(n+\kappa)(n-1+\kappa)}\label{coef2}.
\eeqr
Differentiating (\ref{rec}) with respect to $z_1$ under the assumption that (\ref{formula1}) is valid when the second quantum number is lower than $n$,  and applying the recurrence relation (\ref{rec}) to get rid of the remaining $z_2$ factors, we obtain:
\beqr
\frac{\partial P_{m,n}^\kappa}{\partial z_1}&=& m\, P_{m-1,n}^{\kappa+1}\nonumber\\
&+&\left[A_{m,n-1}(\kappa)\, \tilde{a}_{m-2,n-2}(\kappa+1)-A_{m-1,n-1}(\kappa)\, \tilde{a}_{m,n-1}(\kappa)\right] P_{m-3,n-2}^{\kappa+1}\nonumber\\
&+&\left[A_{m,n-1}(\kappa)+m\, \tilde{a}_{m-1,n-1}(\kappa+1)-(m-1)\, \tilde{a}_{m,n-1}(\kappa)\right] P_{m-2,n-1}^{\kappa+1}\nonumber\\
&+&\left[B_{m,n-1}(\kappa)-(m+1)\, c_{n-1}(\kappa)+m\, c_{n-1}(\kappa+1)\right] P_{m,n-2}^{\kappa+1}\nonumber\\&+&\left[B_{m,n-1}(\kappa)\, c_{n-3}(\kappa+1)-B_{m+1,n-2}(\kappa)\, c_{n-1}(\kappa)\right] P_{m+1,n-4}^{\kappa+1}\nonumber\\
&+&[-\tilde{a}_{m,n-1}(\kappa)\, B_{m-1,n-1}(\kappa)+\tilde{a}_{m,n-3}(\kappa+1)\, B_{m,n-1}(\kappa)\nonumber\\&+&A_{m,n-1}(\kappa)\, c_{n-2}(\kappa+1)-A_{m+1,n-2}(\kappa)\, c_{n-1}(\kappa)] P_{m-1,n-3}^{\kappa+1}  
\eeqr
and by explicit use of (\ref{coef1}) and (\ref{coef2}), we find:
\beqr
A_{m,n-1}(\kappa)\, \tilde{a}_{m-2,n-2}(\kappa+1)-A_{m-1,n-1}(\kappa)\, \tilde{a}_{m,n-1}(\kappa)&=&0\nonumber\\ 
B_{m,n-1}(\kappa)\, c_{n-3}(\kappa+1)-B_{m+1,n-2}(\kappa)\, c_{n-1}(\kappa)&=&0\nonumber\\
-\tilde{a}_{m,n-1}(\kappa)\, B_{m-1,n-1}(\kappa)+\tilde{a}_{m,n-3}(\kappa+1)\, B_{m,n-1}(\kappa)& &\nonumber\\+A_{m,n-1}(\kappa)\, c_{n-2}(\kappa+1)-A_{m+1,n-2}(\kappa)\, c_{n-1}(\kappa)&=&0
\eeqr
and
\beqr
A_{m,n-1}(\kappa)+m\, \tilde{a}_{m-1,n-1}(\kappa+1)-(m-1)\, \tilde{a}_{m,n-1}(\kappa)&=&A_{m,n}(\kappa)\nonumber\\
B_{m,n-1}(\kappa)-(m+1)\, c_{n-1}(\kappa)+m\, c_{n-1}(\kappa+1)&=&B_{m,n}(\kappa)
\eeqr
which establishes the desired result. The proof of (\ref{formula2}) takes advantage of the twin recurrence relation to (\ref{rec}), see \cite{pe99}, 
and is completely analogous. In conclusion we would like to mention that the approach of this note may be used also for the $A_n$ case. We hope to return to this problem in the future.\\\\\\
{\Large\bf Acnowledgments}\\\\
We are grateful to Prof. M. Lorente for interesting discussions. One of the authors (A. M. P.) would like to express his gratitude to the Department of Physics of the University of Oviedo for the hospitality during  his stay as a Visiting Professor.\\

\end{document}